\documentclass[prl,twocolumn,showpacs,amsmath,amssymb,superscriptaddress]{revtex4-1}
\usepackage{graphicx}
\usepackage{dcolumn}
\usepackage{bm}
\usepackage{bbm}
\usepackage{ulem}
\usepackage{color}
\usepackage{slashed}
\usepackage{hyperref}
\usepackage{mathrsfs}
\usepackage{subfigure}
\usepackage{verbatim}
\usepackage{dsfont}
\usepackage{float}

\newcommand{\nc}{\newcommand}
\nc{\be}{\begin{equation}}
\nc{\ee}{\end{equation}}
\nc{\bea}{\begin{eqnarray}} \nc{\eea}{\end{eqnarray}}
\nc{\dg}{\dagger}
\nc{\ua}{\uparrow} \nc{\da}{\downarrow}

\begin{document}

\bibliographystyle{apsrev4-1}

\title{Electrically tunable spin polarization of chiral edge modes in a quantum anomalous Hall insulator}
\author{Rui-Xing Zhang}
\affiliation{Department of Physics, The Pennsylvania State University, University Park, Pennsylvania 16802}
\author{Hsiu-Chuan Hsu}
\affiliation{Department of Physics, The Pennsylvania State University, University Park, Pennsylvania 16802}
\author{Chao-Xing Liu}
\email{cxl56@psu.edu}
\affiliation{Department of Physics, The Pennsylvania State University, University Park, Pennsylvania 16802}
\date{\today}

\begin{abstract}
In the quantum anomalous Hall effect, chiral edge modes are expected to conduct spin polarized current without dissipation and thus hold great promise for future electronics and spintronics with low energy consumption. However, spin polarization of chiral edge modes has never been established in experiments. In this work, we theoretically study spin polarization of chiral edge modes in the quantum anomalous Hall effect, based on both the effective model and more realistic tight-binding model constructed from the first principles calculations. We find that spin polarization can be manipulated by tuning either a local gate voltage or the Fermi energy. We also propose to extract spin information of chiral edge modes by contacting the quantum anomalous Hall insulator to a ferromagnetic (FM) lead. The establishment of spin polarization of chiral edge modes, as well as the manipulation and detection in a fully electrical manner, will pave the way to the applications of the quantum anomalous Hall effect in spintronics.
\end{abstract}

\pacs{73.20.-r, 72.25.-b, 85.75.-d}

\maketitle

{\it Introduction} -
For a two dimensional electron gas under a strong magnetic field, Landau levels are formed and drive the system into a state characterized by the zero longitudinal resistance and the quantized Hall conductance with an integer value of $\frac{e^2}{h}$. This phenomenon is known as the quantum Hall (QH) effect, which was discovered by Von. Klitzing in 1980 \cite{klitzing1980}. Recently, it was theoretically predicted \cite{haldane1988,liu2008,yu2010} that this type of quantization in the Hall conductance can also be realized in magnetic insulating materials at a zero external magnetic field. This phenomenon, dubbed as the ``quantum anomalous Hall (QAH) effect'', was soon observed experimentally in magnetically doped topological insulators (TIs), the Cr or V doped (Bi,Sb)$_2$Te$_3$ films \cite{chang2013,kou2014,checkelsky2014,bestwick2015,chang2015}. The physical origin of the QAH effect is spin-orbit coupling and exchange coupling between magnetic moments and electron spins in magnetic materials, rather than magnetic fields and the associated Landau levels \cite{haldane1988}.
The experimental observation of the exact quantization of Hall conductance and neglegible longitudinal resistance confirm the dissipationless nature of transport for the QAH effect \cite{bestwick2015,chang2015} and the mapping of global phase diagram establishes the topological equivalence between the QH effect and the QAH effect \cite{checkelsky2014,kou2015}.

Similar to the case of the QH effect, dissipationless currents in the QAH insulators are carried by one dimensional (1D) chiral edge modes (CEMs), which propagate along one direction at the edge of the system and are robust against disorder scatterings. CEMs are believed to hold great promise for the potential applications in electronics and spintronics with low power consumption \cite{zhang2012}. For any spintronic application, it is required for CEMs to carry spin polarization (SP). Naively, one may expect that SP of CEMs should exist and follows bulk magnetization, but this issue has seldom been studied theoretically.
In Ref. \cite{wu2014}, SP of CEMs was studied in the context of a two band model, which is more relevant to cold atom systems. For condensed matter systems, it is more complicated since spin and orbital are mixed with each other due to spin-orbit coupling \cite{supplementary}.


In this letter, we investigate SP of CEMs of the QAH effect in magnetically doped (Bi,Sb)$_2$Te$_3$. Surprisingly, we find that SP of CEMs though exists but does not follow bulk magnetization, and sensitively depends on the boundary conditions. In particular, we find that the direction of SP of CEMs can be manipulated by a local gate voltage, thus opening the possibility of controlling SP of CEMs in a fully electrical manner. We provide a simple physical picture of SP of CEMs based on the effective four-band model and further study its behavior in the more realistic calculations based on the tight-binding model constructed by the Wannier function method of the first principles calculations \cite{marzari1997,souza2001}. We propose the spin valve effect \cite{dieny1994} of CEMs in a standard experimental setup of the Hall measurement with ferromagnetic (FM) leads to extract spin information of CEMs.

{\it Spin polarization of chiral edge modes} -
To study SP of CEMs in a QAH insulator, we first consider an effective four band model for magnetically doped (Bi,Sb)$_2$Te$_3$ films \cite{yu2010}. The low energy physics of this system is dominated by two surface states from top and bottom surfaces, which hybridize with each other to open a gap due to the finite size effect. Meanwhile, the exchange coupling of electron spin arises from the doping of magnetic atoms.  Thus, the effective Hamiltonian $H_{eff}$ is given by
\begin{eqnarray}
	H_{eff}&=&v\tau_z\otimes(k_y\sigma_x-k_x\sigma_y)+m({\bf k})\tau_x\otimes\sigma_0+{\bf M}\cdot \tau_0\otimes\vec{\sigma} \nonumber\\
&&+V(x,y)\tau_z\otimes\sigma_0+A(k)\tau_0\otimes\sigma_0 \nonumber \\
&&+h(k_+^3+k_-^3)\tau_0\otimes\sigma_z,
\label{Eq:effective model}
\end{eqnarray}
where $\sigma_{x,y,z}$ and $\tau_{x,y,z}$ matrices are Pauli matrices of spin and layer (top or bottom) degree of freedom, and $\sigma_0=\tau_0=I_{2\times2}$. $m({\bf k})=m_0+m_2(k_x^2+k_y^2)$ gives hybridization between the top and bottom surface states. The $\bf{M}$ term describes exchange coupling between electron spin and magnetization of magnetic doping. We only consider the out-of-plane magnetization $M_z$. $A(k)=A_0+A_1(k_x^2+k_y^2)$ is a higher order correction of the surface states and $h$ is defined as the strength of hexagon warping effect \cite{liu2010,fu2009}. The $V$ term denotes the asymmetric potential between the top and bottom layers which has spatial dependence and can be induced by a global or local gate voltage.

\begin{figure}[t]
\centering
\includegraphics[width=3.4 in]{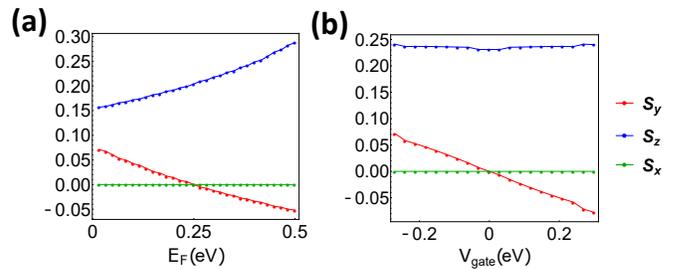}
\caption{Spin polarization $S_{x,y,z}$ of chiral edge state in our effective model is plotted at fixed local gate voltage ($V=0.1$ eV) in (a), and at fixed Fermi energy ($E_F=0.35$ eV) in (b).}
\label{Figure:effmodel}
\end{figure}

The edge spectrum of our effective model, as well as the corresponding SP, can be evaluated with the help of the iterative Green's function method \cite{sancho1984,sancho1985}. We consider a semi-infinite system with the $x$ direction still translationally invariant and one edge parallel to $x$ direction (x-edge). The following set of parameters, $v=1,M_z=0.5,m_0 = 0.1,m_2 = 0.25,A_0=0.15,A_1=0.05,h=0.1$, is chosen to keep the system in the QAH regime. The local Green's function $G(k_x,\omega)$ on the x-edge can be evaluated iteratively. The total local density of states (DOS) and the spin DOS along the $i$th direction ($i=x,y,z$) are defined as $\rho_0=-\frac{1}{\pi}\text{Im} [tr(G)]$ and $\rho_{\sigma_i}=-\frac{1}{\pi}\text{Im} [tr(G\sigma_i)]$, respectively. As shown in the Supplementary Materials \cite{supplementary}, the CEM can be easily identified from the peak of local DOS $\rho_0$ on the x-edge. The corresponding SP $S_i$ along the $i$th direction ($i=x,y,z$) for the CEM is normalized by the total DOS as $S_i=\rho_{\sigma_i}/\rho_0$.
In Fig. \ref{Figure:effmodel} (a) and (b), the SP of the CEMs is plotted for all three spin components with different local gate voltages and chemical potentials. Let us take the lattice constant to be unity and label the lattice site with an integer index $n\geq 1$, with $(n-1)$ being the distance between the $n$th lattice site and the boundary. Here the local gate voltage $V$ is added only on all $n=1$ lattice sites, which leaves the bulk states unchanged. We find that for a zero gate voltage ($V=0$), only the $z$ component ($S_z$) SP of CEMs is non-zero. If we apply a local gate voltage, SP becomes non-zero for both the $y$ and $z$ direction, but still keeps zero for the $x$ direction. Therefore, SP of CEMs can exist within the $y$-$z$ plane and can be controlled by a local gate voltage.
It is also interesting to notice that the local gate voltage mainly controls the amplitude of $S_y$, but barely change that of $S_z$ (see Fig. \ref{Figure:effmodel}(b))\cite{supplementary}. The chemical potential can also tune the magnitude of $S_z$ (both $S_y$ and $S_z$) at a zero (finite) local gate voltage, as shown in Fig. \ref{Figure:effmodel} (a). Therefore, our effective model (Eq. \ref{Eq:effective model}) suggests that SP of CEMs can be manipulated by either applying a local gate voltage or tuning the chemical potential.

Next we provide an analytical solution of the eigen wave function for the Hamiltonian (Eq. \ref{Eq:effective model}) with $V=A_0=A_1=h=0$ to understand the electrical tunability of SP of CEMs. Assume $m_{0,2}>0$, the system will enter QAHE regime when $|M_z|>m_0$. The zero mode of the Hamiltonian $H_{eff}$ localizing on the edge can be solved exactly \cite{supplementary} as
\be
\Psi(y)=C(e^{-\lambda_+^- y}-e^{-\lambda_-^- y})[|t\rangle\otimes(|\ua_y\rangle)+|b\rangle\otimes(|\da_y\rangle)],
\label{Eq:effspin}
\ee
where $C$ is a normalization factor and $\lambda_{\pm}=\frac{1}{2m_2}(v\pm\sqrt{v^2+4m_2(m_0-M_z)})$. Here $|\ua_y (\da_y)\rangle$ denote spin up (down) state along the $y$ direction, and $|t(b)\rangle$ denotes the contribution from the top (bottom) layer of thin films. The wave function (\ref{Eq:effspin}) of CEMs consists of two parts: one part on the top surface with SP along the $+y$ direction and the other on the bottom surface with SP along the $-y$ direction. Thus, the SP is locked to the layer (top or bottom) for the CEMs. A local gate voltage can push the wave functions of CEMs into one layer, thus leading to the change of SP.

\begin{figure}[t]
\centering
\includegraphics[width=3.3 in]{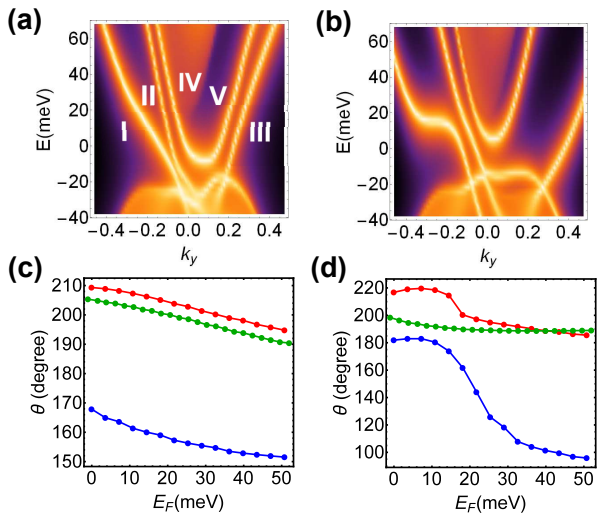}
\caption{Edge spectrum and edge state SP angle $\theta$ for magnetically doped Sb$_2$Te$_3$ QAH systems are plotted with: (1) $V=0$ eV in (a) and (c). (2) $V=0.1$ eV in (b) and (d). In (c) and (d), SP angle of edge mode I, II and III are plotted in red, blue and green.}
\label{Figure:sb2te3}
\end{figure}

The above analysis of SP of CEMs is based on the effective four-band model (Eq. \ref{Eq:effective model}) and one may ask if these results still hold for a realistic system, such as Cr or V doped (Bi,Sb)$_2$Te$_3$. To answer this question, we carry out explicit calculations for a magnetically doped Sb$_2$Te$_3$ thin film system with the realistic tight-binding model constructed from the maximal localized Wannier function (MLWF) method \cite{marzari1997,souza2001} based on the first principles calculations, which has been successfully applied to the quantitative study of the QAH effect in magnetically doped (Bi,Sb)$_2$Te$_3$ \cite{wang2013a,wang2013b}. The exchange coupling between electron spin and magnetization is included in the tight-binding model in the mean field approximation $H_{ex}=\lambda_{Sb}\sigma_z^{Sb}+\lambda_{Te}\sigma_z^{Te}$. Here we consider only the contribution from out-of-plane magnetization and $\sigma_z^{Sb(Te)}$ is the z directional spin operator for Sb (Te) atoms.
In our calculation, we consider a film with two quintuple layers and choose the exchange coupling strength to be $\lambda_{Sb}=0.4$ eV and $\lambda_{Te}=0.0$ eV, which is strong enough to drive the system into the QAH phase. The edge dispersion is also calculated with the iterative Green function method in a semi-infinite configuration along (100) direction, as shown in Fig. \ref{Figure:sb2te3} (a). For edge modes, we find three left movers (the modes I, II and IV) and two right movers (the modes III and V), suggesting that one chiral edge mode (left mover) and the other two trivial 1D edge modes (or non-chiral edge modes) \cite{wang2013a}. Since the mode IV is directly connected to V, these two modes must be trivial 1D edge modes and thus we focus on the modes I, II and III below. The influence of the local gate voltage ($V=0.1$ eV) is shown in Fig. \ref{Figure:sb2te3} (b), from which one can see that the number of left and right movers is unchanged, but their energy dispersions are shifted.

After obtaining the edge state Green function, it is straightforward to calculate the corresponding SP vector ${\bf S}$. First of all, $S_x$ is vanishingly small compared to other spin components. This confirms the results from the effective four band model. Thus, ${\bf S}$ only appears in the $y$-$z$ plane and can be characterized by its magnitude $|S|$ and a polarization angle $\theta$ relative to $+z$ axis.
For $S_y$, we notice that it vanishes at $V=0$ for the effective four band model, while it is non-zero in the realistic tight-binding model. This difference is because of the out-of-plane mirror symmetry in our four band model, which is absent in realistic crystals. Therefore, both $S_y$ and $S_z$ are non-zero even at $V=0$ in realistic tight-binding model. We plot Fermi energy $E_F$ dependence of polarization angle $\theta$ at the local gate $V=0$ eV and $V=0.1$ eV in Fig. \ref{Figure:sb2te3} (c) and (d).
The polarization angles $\theta$ for the modes I (red lines) and III (green lines) behave similarly to each other, while that of the mode II (blue lines) reveals a completely different characteristics. This is a clear evidence that the modes I and III are connected to each other, forming a non-chiral edge mode, while the mode II can be identified as the non-trivial CEM. In Fig. \ref{Figure:sb2te3} (d), we notice an abrupt change of polarization angle $\theta$ for the modes I and II when $E_F$ is between 0 meV and 20 meV. Compared with the energy dispersion in Fig. \ref{Figure:sb2te3} (b), we find that this change results from the strong hybridization effect between the modes I and II. Thus, the CEM and non-chiral mode are not well defined in this regime. In the other regime, we find a smooth change of SPs with respect to local gate voltages and chemical potentials.
{\it Experimental detection of edge spin polarization} -
Our theoretical calculations based on both the effective model and realistic tight-binding model have clearly demonstrate electrically tunable SP in magnetically doped (Bi,Sb)$_2$Te$_3$. However, the experimental detection of SP is not an easy task since the bulk is FM and it is unclear how to distinguish SP of CEMs from that of bulk ferromagnetism by magnetization measurement. In contrast, when the Fermi energy is tuned into the bulk gap, the transport signals are dominated by CEMs. Thus, it is more promising to search for SP of CEMs from transport measurements. 

\begin{figure}[t]
\centering
\includegraphics[width=3.4 in]{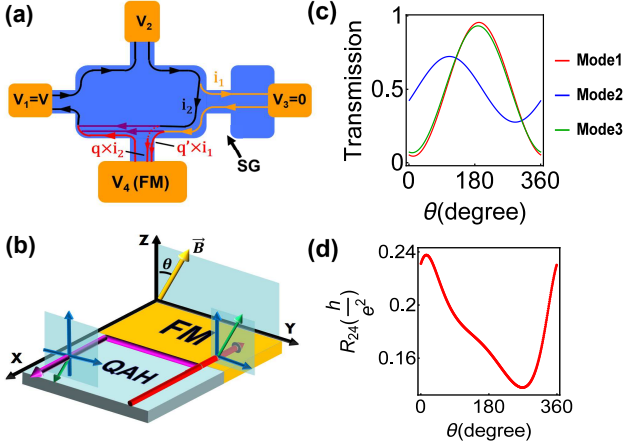}
\caption{Our proposed four terminal transport setup is shown in (a) with lines showing current flows. In (b), we show the spin valve effect at the FM lead $V_4$: Only current with spin parallel to ${\bf M}$ can flow into $V_4$, while those with spin anti-parallel to ${\bf M}$ will be reflected. $\theta$ dependence of transmission $q_i$s at $E_F=33$ meV are plotted in (c). Based on this, we plot the evolution of transverse resistance $R_{24}$ with $\theta$ in (d). }
\label{Figure:setup}
\end{figure}

Our proposal is based on a four terminal device with a FM probe as the fourth probe, as shown in Fig. \ref{Figure:setup} (a). This experimental setup is similar to that in the study of disordered leads in the QH system and here FM leads play the role of disordered leads \cite{datta1997}. When SP of the CEM is parallel to magnetization ${\bf M}$ of the FM lead, it can flow into the lead, while when they are opposite, it will be scattered. We apply a voltage drop between the leads $V_1$ and $V_3$ ($V_1=V$ and $V_3=0$) and also introduce a split gate, denoted as SG in the Fig.\ref{Figure:setup} (a). Due to the split gate, there are two types of currents flowing from the lead $V_2$ to $V_4$: the current $i_1$, which goes through $V_3$, and the current $i_2$ flowing directly from $V_2$ to $V_4$. Based on the Landauer-B$\ddot{\text u}$ttiker formula \cite{buttiker1986,buttiker1988}, the current $i_1$ to the lead $V_4$ shares the same chemical potential as $V_3=0$, while the current $i_2$ is determined by the chemical potential in the lead $V_2=V_1=V$. Importantly, we assume that chemical potentials of two currents $i_1$ and $i_2$ have not reached equilibrium when they enter the FM lead. This is determined by the inelastic scattering length, which is estimated as several $\mu m$ for the QH case \cite{koch1991}, and we expect a similar length scale in our case. Thus, the corresponding SPs are also expected to be different for these two currents. Since the scattering rate into the FM lead $V_4$ depends on the relative angle between the SP of CEMs and ${\bf M}$. We expect the transmissions of $V_2 \rightarrow V_4$ and $V_3 \rightarrow V_4$ also depend on ${\bf M}$ of the FM lead. As a result, the chemical potential in $V_4$ will vary when rotating ${\bf M}$. This provides a detectable signal, which is similar to the spin valve effect \cite{dieny1994}, in transport measurements and can be directly related to SP of CEMs. It should be emphasized that the split gate SG plays an essential role here. Without the split gate, all currents flowing into $V_4$ come from $V_3$, and thus the corresponding chemical potential in $V_4$ must be equal to that of $V_3$, independent of magnetization direction of the FM lead.

To formulate this idea, we assume that only p ($p<1$) fraction of the net edge current $i_0$ can flow from $V_2$ to $V_3$ due to the split gate, as shown in Fig. \ref{Figure:sb2te3} (a), so $i_1=pi_0, i_2=(1-p)i_0$. We further assume the $q'$ ($q$) fraction of the current $i_1$ ($i_2$) has spin polarizing along ${\bf M}$. Since SP of CEMs depends on chemical potential, it is reasonable to assume $q\neq q'$.
Therefore, the currents from $V_2$ ($V_3$) to $V_4$ are $q'i_1$ and $qi_2$, respectively, as shown in Fig. \ref{Figure:setup} (a). The Hall conductance $G_{24}$ can be derived based on Landauer-B$\ddot{\text u}$ttiker formula \cite{supplementary} as
\bea
G_{24}=\frac{I_1}{V_2-V_4}=G_0\frac{(1-p)q+pq'}{q'}
\label{Eq:conductance_one mode}
\eea
with the conductance quanta $G_0=\frac{e^2}{h}$.
From Eq. (\ref{Eq:conductance_one mode}), one can see that when there is no split gate ($p=1$), $G_{24}=G_0=\frac{e^2}{h}$, which recovers the quantized Hall conductance and is independent of its spin information. When a split gate is introduced, $0<p<1$ and $q\neq q'$, and $G_{24}$ will deviate from the quantized value and we discuss how to extract the information of SP of CEMs from the Hall resistance measurement in the Supplementary Materials \cite{supplementary}.

For the realistic systems of magnetically doped (Bi,Sb)$_2$Te$_3$, we have shown additional non-chiral modes coexisting with CEMs.
Thus, it is natural to ask how these non-chiral modes influence the above transport measurement.
We consider one pair of non-chiral edge mode (mode I and III in Fig. \ref{Figure:sb2te3} (a)) and one CEM (mode II) and further assume no inter-channel scattering between different modes for simplicity. We take the modes I and II flowing clockwise (as shown in Fig. \ref{Figure:setup} (a)), and the mode III flowing counter-clockwise (flipping the directions of currents in Fig. \ref{Figure:setup} (a)). $q_{1,2,3}$ ($q'_{1,2,3}$) is defined as the transmission into FM probe of edge current $i_2$ ($i_1$) for the modes I,II and III, with $Q=q_1+q_2,Q'=q_1'+q_2'$ and $\bar{p}=1-p$. The explicit expression of transverse resistance is shown in the Supplementary Materials \cite{supplementary}. In the limit where the split gate vanishes ($p=1-\bar{p}=1$), the transverse conductance $G_{24}$ becomes
\be
G_{24}=\frac{e^2}{h}\frac{7(Q'+q_3)-3q_3Q'}{2Q'-q_3}
\label{Eq:cond_sb2te3}
\ee
which deviates from the quantized value. Thus, in contrast to the single CME case, Hall resistance will depend on the magnetization direction of FM leads even without any split gate for the case with both CME and non-chiral modes.


To apply Eq. (\ref{Eq:cond_sb2te3}) to magnetically doped Sb$_2$Te$_3$ films, we need to extract the coefficients $q_i$s from our realistic tight-binding model. As discussed above, $q_i$s are determined by the projection of SP of CEMs into the magnetization direction ${\bf M}$ of FM leads. Since SP of CEMs only exists in the $y$-$z$ plane, we only concern the projection of SP into the $y$-$z$ plane. Let's assume ${\bf M}$ has an angle $\theta$ relative to $+z$ direction in the $y$-$z$ plane, as shown in Fig. \ref{Figure:setup} (b). The corresponding projection operator for CEMs is defined as:
\be
P^{\ua\ua}_{\theta}=|\ua_{\theta}\rangle\langle \ua_{\theta}|, \text{ with }|\ua_{\theta}\rangle=e^{-i\sigma_x\frac{\theta}{2}}|\ua_z\rangle
\ee
Consequently, the angle dependent transmission $q_i(\theta)$ for the $i$th edge mode is given by
\be
q_i(\theta)
=\frac{\langle \psi_i|P^{\ua\ua}_{\theta}|\psi_i \rangle}{\langle \psi_i|P^{\ua\ua}_{\theta}|\psi_i \rangle+\langle \psi_i|P^{\da\da}_{\theta}|\psi_i\rangle}
\label{Eq:transmission}
\ee
where $|\psi_i\rangle$ is the wave function for the mode $i$.
With Eq. \ref{Eq:cond_sb2te3} and Eq. \ref{Eq:transmission}, we can calculate the transmissions $q_i(\theta)$s for the modes I, II and III as a function of $\theta$ and Fermi energy for the local gate voltage $V=0.1 eV$ for our realistic tight-binding model \cite{supplementary}. For the chemical potential $E_F=33$ meV, $\theta$ dependence of transmission $q_i$s and the Hall resistance $R_{24}$, are shown in Fig. \ref{Figure:setup} (c) and (d), respectively.
$R_{24}$ shows a strong dependence on $\theta$, thus revealing the spin valve effect for CEMs \cite{dieny1994}. This provides a very clear and experimentally feasible evidence to detect spin signal in a QAH insulator.

{\it Discussions} -
In summary, we have shown that SP of CEMs can be generated, manipulated and detected in a QAH insulator in a fully electric manner. This paves the way of potential applications for the QAH effect in spintronics. Disorder is inevitable in realistic samples and we show the stability of SP of CEMs against disorder in the Supplementary Materials \cite{supplementary}. Although we focus on the magnetically doped (Bi,Sb)$_2$Te$_3$ films here, the electric controllability of SP of CEMs is a general property of a QAH insulator. The bulk topology (non-zero Chern number) only guarantees the existence of CEMs, but the detailed form of wave functions of CEMs depends on local electric environment, and thus can be controlled by a local gate voltage. We expect a similar effect also occurs in other QAH insulators, such as Mn doped HgTe quantum wells \cite{liu2008} and InAs/GaSb quantum wells \cite{wang2014}, where the local gate voltage can induce a local Rashba spin-orbit coupling and tilt SP of CEMs.

{\it Acknowledgement} -
We would like to acknowledge Cui-Zu Chang, Moses Chan, Abhinav Kandal, Jainendra Jain, Jagadeesh Moodera, Xiao-Liang Qi and Nitin Samarth for the helpful discussions.


\bibliography{spin}

\onecolumngrid
\newpage

\subsection{\large Supplementary Materials for ``Edge spin polarization in Quantum Anomalous Hall Insulators"}

\subsection{Analytical solution of edge wave function in effective model}
To solve for the zero mode wave function of the effective model, we can first apply a unitary transformation $U$,
\be
U=\frac{1}{\sqrt{2}}
\begin{pmatrix}
1 & 0 & 1 & 0 \nonumber \\
0 & 1 & 0 & -1 \nonumber \\
0 & 1 & 0 & 1  \nonumber \\
1 & 0 & -1 & 0
\end{pmatrix},
\ee
leading to
\be
H'=UH_{eff}U^{\dagger}=
\begin{pmatrix}
m_k+M_z & iv k_- & M_- & 0 \nonumber \\
-iv k_+ & -m_k-M_z & 0 & M_+ \nonumber \\
M_+ & 0 & m_k-M_z & -iv k_+  \nonumber \\
0 & M_- & iv k_- & -m_k+M_z
\end{pmatrix}\label{Eq:Ham_app1}
\ee
We set $M_+=M_-=0$ and $H$ becomes block diagonal. We take $m(k)=m_0+m_2k^2$ and assume $m_{0,2}>0$, and band inversion happens when $|M_z|>m_0$. We would like to emphasize that the two-band model taken in Ref. \cite{wu2014} is just one block of the above four-band Hamiltonian (\ref{Eq:Ham_app1}). Nevertheless, the bases for the Hamiltonian (\ref{Eq:Ham_app1}) are $(|+,\ua\rangle,|-,\da\rangle,|+,\da\rangle,|-,\ua\rangle)$, where "$+/-$" denotes bonding/anti-bonding state between top and bottom surface states. The corresponding spin operators are given by
\be
s_x=\begin{pmatrix}
0 & 0 & 1 & 0 \nonumber \\
0 & 0 & 0 & 1 \nonumber \\
1 & 0 & 0 & 0  \nonumber \\
0 & 1 & 0 & 0
\end{pmatrix},
s_y=\begin{pmatrix}
0 & 0 & -i & 0 \nonumber \\
0 & 0 & 0 & i \nonumber \\
i & 0 & 0 & 0  \nonumber \\
0 & -i & 0 & 0
\end{pmatrix},
s_z=\begin{pmatrix}
1 & 0 & 0 & 0 \nonumber \\
0 & -1 & 0 & 0 \nonumber \\
0 & 0 & -1 & 0  \nonumber \\
0 & 0 & 0 & 1
\end{pmatrix}
\ee
Thus, we find that the spin operators in the four band model are different from those used in the two-band model in Ref. \cite{wu2014}, which are more relevant to the cold atom system.

Now we consider the case with $M_z>0$, in which the lower block of $H'$ is in the QAH phase while the upper block is in the trivial phase. For the lower block, we have
\be
H_{lower}=
\begin{pmatrix}
m_k-M_z & -iv k_+  \nonumber \\
iv k_- & -m_k+M_z
\end{pmatrix}
=(m_k-M_z)\sigma_z+v(k_y\sigma_x+k_x\sigma_y).
\ee
We choose the following set of parameters which satisfies the above band inversion condition: $v=1,M_z=0.5,m_0=0.1,m_2=0.25$. 

The wave-function of zero energy mode of the chiral edge mode can be solved for the lower block Hamiltonian.
Let us assume the edge is along the y direction, so we consider $k_x=0$ first, and check the eigen equation for zero mode
\be
[(m_0-m_2\partial^2_y-M_z)\sigma_z-iv_x\sigma_x\partial_y]\psi(y)=0
\ee
Multiply both sides by $\sigma_x$,
\be
[(m_0-m_2\partial^2_y-M_z)\sigma_y+v_x\partial_y]\psi(y)=0
\ee
Consider the ansatz $\psi(y)=Ce^{-\lambda y}\xi_s$, where $\sigma_y\xi_s=s\xi_s,s=\pm1$, we have
\be
\lambda_{\pm}^s=\frac{1}{2m_2}(-sv\pm\sqrt{v^2+4m_2(m_0-M_z)})
\ee
Thus the wave function should be the following linear combination:
\be
\Psi(y)=\sum_s (c_{s+}e^{-\lambda_{+}^s y}+c_{s-}e^{-\lambda_{-}^s y})\xi_s
\ee
with
\bea
&\lambda_{+}^s+\lambda_{-}^s&=-\frac{sv}{m_2} \nonumber \\
&\lambda_{+}^s\lambda_{-}^s&=-\frac{m_0-M_z}{m_2}
\eea
Assuming the material (vacuum) region is at $x>0$ ($x<0$), then we require that wave function vanishes at both $x=+\infty$ and $x=0$. This requires $c_{s+}=-c_{s-}$, $\lambda_{\pm}^s>0$ and $s$ to be negative. So the wave function is given by
\be
\Psi(y)=c(e^{-\lambda_+^- y}-e^{-\lambda_-^- y})\xi_-
\ee
Notice that the spinor part $\xi_-$ is defined in the hybridized bases $(|+\da\rangle,|-\ua\rangle)$. Then $\xi_-$ can be written as
\bea
\xi_{-}&\sim& i|+\da\rangle+ |-\ua\rangle=i(|t\da\rangle+|b\da\rangle)+(|t\ua\rangle-|b\ua\rangle) \nonumber \\
&=&|t\rangle\otimes(i|\da\rangle+ |\ua\rangle)-|b\rangle\otimes(-i|\da\rangle+|\ua\rangle)  \nonumber \\
&=&i|t\rangle\otimes(|\ua_y\rangle)+i|b\rangle\otimes(|\da_y\rangle)
\label{Eq:spinor}
\eea
We emphasize again that the above solution does not correspond to a net spin polarization along the y direction, as discussed in Ref. \cite{wu2014}, due to the different bases. The wave function of CEMs consists of two parts: one part on the top surface with SP along the $+y$ direction and the other on the bottom surface with SP along the $-y$ direction. These two contributions exactly cancel each other, leading to zero net SP of CEMs. From the symmetry perspective, we notice that there is an emergent mirror symmetry between the top and bottom surfaces, which requires $S_y=0$. Therefore, applying a local gate voltage breaks this mirror symmetry and yields an imbalance of SP between the top and bottom surfaces, giving rise to a finite $S_y$. The absence of $S_z$ is completely accidental, as a result of setting $A_{0,1}=h=0$. Numerically, when assigning non-zero values to $A_{0,1}$ and $h$, non-zero $S_z$ appears, as shown in Fig. [1] in the main text. Therefore, $S_z$ is insensitive to this mirror symmetry breaking, just as we have numerically verified. To conclude, applying a local gate voltage changes both magnitude and direction of CEM SP.

\section{Transverse conductance from Landauer-B$\ddot{\text u}$ttiker formula}
\begin{figure}[t]
\centering
\includegraphics[width=0.9 \textwidth]{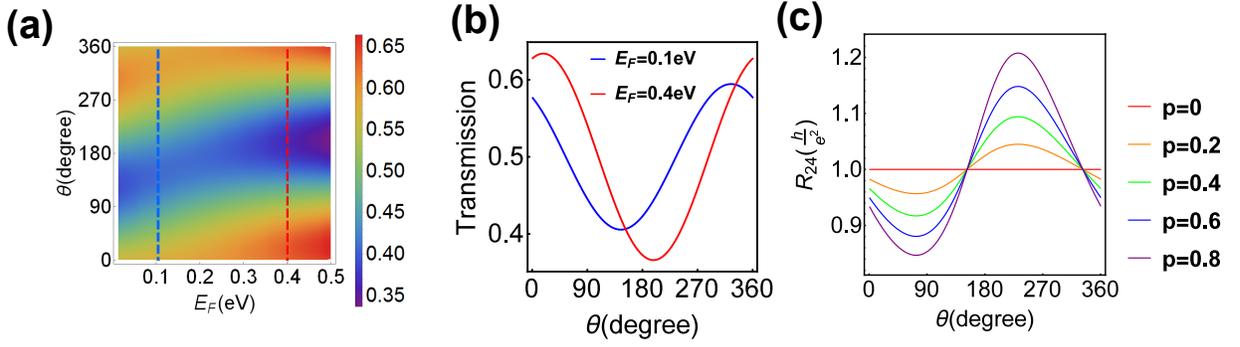}
\caption{In (a), we plot $\theta$ dependence of transmission $q$ at various chemical potential for the chiral edge state in the effective model. In (b), we fix the chemical potential and plot the $\theta$ dependence of $q$. Transverse resistance $R_{24}$ at different split gate transmission $p$ is plotted in (c).}
\label{Figure:conductance_effective}
\end{figure}

In this section, we will derive analytical expressions of transverse conductance based on Landauer-B$\ddot{\text u}$ttiker formula \cite{buttiker1986,buttiker1988}. We will first discuss the situation with one single chiral edge mode, and then generalize the expression to a case where a chiral edge mode and one pair of non-chiral edge mode coexist.

Following the setup configuration in Fig. [3] (a) in the main text, the conductance matrix G for a single chiral edge state (flowing clockwise) is given by
\be
G=G_0\times
\begin{pmatrix}
0 & (1-p)(1-q) & p(1-q') & (1-p)q+pq' \nonumber \\
1 & 0 & 0 & 0 \nonumber \\
0 & p & 0 & 0 \nonumber \\
0 & (1-p)q & pq' & 0
\end{pmatrix}
\label{Eq:G}
\ee
where the bases for matrix G is ($V_1,V_2,V_3,V_4$). Here $p$ is the split gate transmission, while $q'$ and $q$ are transmissions into FM probe $V_4$ of currents from $V_2$ and $V_3$. Here a matrix element $G_{ij}$ 
describes the conductance contribution from probe $V_j$ to probe $V_i$. Then, current voltage relations are
\be
\begin{pmatrix}
I_1 \nonumber \\
I_2 \nonumber \\
I_3 \nonumber \\
I_4 \nonumber \\
\end{pmatrix}
=G_0\times
\begin{pmatrix}
1 & -(1-p)(1-q) & -p(1-q') & -(1-p)q-pq' \nonumber \\
-1 & 1 & 0 & 0 \nonumber \\
0 & -p & p & 0 \nonumber \\
0 & -(1-p)q & -pq' & (1-p)q+pq'
\end{pmatrix}
\begin{pmatrix}
V_1 \nonumber \\
V_2 \nonumber \\
V_3 \nonumber \\
V_4 \nonumber \\
\end{pmatrix}
\label{Eq:landauer}
\ee
Here $I_i$ ($i\in {1,2,3,4}$) denotes the current that flows into the $i$th probe. Notice that $I_2=I_4=V_3=0$ and $I_1=-I_3$, giving rise to
\bea
V_1&=&V_2 \nonumber \\
(1-p)qV_2&=& [(1-p)q+pq']V_4 \nonumber \\
I_1&=&G_0pV_1
\eea
The transverse conductance $G_{24}^{\text{chiral}}$ is given by
\be
G_{24}^{\text{chiral}}=\frac{I_1}{V_2-V_4}=G_0\frac{(1-p)q+pq'}{q'}
\label{Eq:conductance_onemode_supplementary}
\ee
Numerically, we plot the $\theta$ dependence of transverse resistance $R_{24}$ in Fig. \ref{Figure:conductance_effective} (c). Here, we take current $i_1$ (from $V_3$) at $E_F=0.1$ eV and $i_2$ (from $V_2$) at $E_F=0.4$ eV. Transmission of $i_1$ and $i_2$ into $V_4$ is calculated in Fig. \ref{Figure:conductance_effective} (a), which will be discussed in details in a later section. In the Fig. \ref{Figure:conductance_effective} (b), we extract the transimission information for $E_F=0.1$eV (blue line) and $E_F=0.4$eV (red line) from Fig. \ref{Figure:conductance_effective} (a), which provides us with the $q$ and $q'$ at different $\theta$. Transmission into the split gate $p$ is chosen at various values. As long as the split gate structure exists ($p\neq0$), there will be an obvious $\theta$ dependence of Hall resistance $R_{24}$, as shown in Fig. \ref{Figure:conductance_effective} (c). But when the split gate structure is absent, transverse resistance will be quantized despite the value of $\theta$. 

For the magnetically doped Sb$_2$Te$_3$ QAH system, we consider two clock-wise modes (mode I and III) and one counter-clockwise mode (mode II). Let us define $q_{1,2,3}$ ($q'_{1,2,3}$) to be the transmission into FM probe of edge current $i_2$ ($i_1$) for the modes I,II and III, with $Q=q_1+q_2,Q'=q_1'+q_2'$ and $\bar{p}=1-p$. So the conductance matrix for mode I and mode II can simply take the form of Eq \ref{Eq:G}, while conductance matrix for mode III is as follows:
\bea
G_I&=&G_0\times
\begin{pmatrix}
0 & (1-p)(1-q_1) & p(1-q'_1) & (1-p)q_1+pq_1' \nonumber \\
1 & 0 & 0 & 0 \nonumber \\
0 & p & 0 & 0 \nonumber \\
0 & (1-p)q_1 & pq_1' & 0
\end{pmatrix} \nonumber \\
G_{II}&=&G_1 (q_1\Leftrightarrow q_2, q_1'\Leftrightarrow q_2') \nonumber \\
G_{III}&=&G_0\times
\begin{pmatrix}
0 & 1 & 0 & 0 \nonumber \\
(1-p)(1-q_3) & 0 & p & (1-p)q_3 \nonumber \\
p(1-q_3) & 0 & 0 & pq_3 \nonumber \\
q_3 & 0 & 0 & 0
\end{pmatrix}
\eea
The net conductance matrix is simply an addition of all three conductance matrix $G_{\text net}=G_I+G_{II}+G_{III}$. After some calculations, we arrive at the following Hall conductance
\be
G_{24}^{\text{non-chiral}}=\frac{e^2}{h}\frac{(7+2\bar{p})[Q'+\bar{p}(Q-Q')]+q_3[7+2\bar{p}-3Q'+\bar{p}(2\bar{p}+1)(Q'-Q)]}{(2+\bar{p})Q'-q_3(1+\bar{p}Q')}
\label{Eq:cond_sb2te3_split}
\ee
In the limit where the split gate vanishes ($p=1-\bar{p}=1$), the transverse conductance $G_{24}^{\text{non-chiral}}$ becomes
\be
G_{24}^{\text{non-chiral}}=\frac{e^2}{h}\frac{7(Q'+q_3)-3q_3Q'}{2Q'-q_3}
\ee

\section{Electron current transmission at the ferromagnetic probe}
In the previous discussion, we have shown that transverse conductance $G_{24}$ has a strong dependence on transmissions at the FM probe $V_4$. In the main text, we have defined the transmission $q_i$ based on a projection operator approach. Since the edge state Green function can be calculated with the help of iterative Green function method, we can easily rewrite the transmission expression in terms of the edge state Green function $G_i(\omega)$ for edge mode i (i=I,II,III)£º
\be
q_i(\theta)=\frac{Im[tr(G_i(\omega)e^{i\frac{\theta}{2}\sigma_x})]}{Im[tr(G_i(\omega)e^{i\frac{\theta}{2}\sigma_x})]
+Im[tr(G_i(\omega)e^{i\frac{\theta+\pi}{2}\sigma_x})]}
\label{Eq:transmission_Green}
\ee

Based on Eq. \ref{Eq:transmission_Green}, in Fig. \ref{Figure:conductance_effective} (a), we calculate the chemical potential and $\theta$ dependence of transmission for the chiral edge state in the effective QAH model. Here we apply a local gate voltage $V=0.3$ eV. The blue line ($E_F=0.1$ eV) and red line ($E_F=0.3$ eV) denote two different chemical potentials due to the split gate effect. The two Fermi energies are the $E_F$s we choose to calculate transverse resistance in Fig \ref{Figure:conductance_effective} (c). We further plot the transmissions $q(E_F)$ at these two fixed Fermi energies in Fig. \ref{Figure:conductance_effective} (a).

\begin{figure}[t]
\centering
\includegraphics[width=0.6 \textwidth]{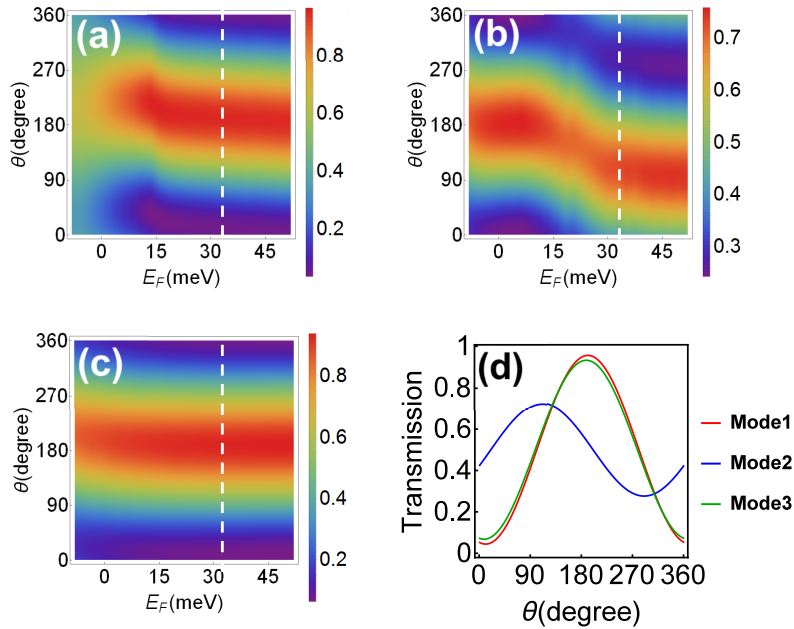}
\caption{Chemical potential and $\theta$ dependence of transmissions of mode I, II and III are plotted in (a), (b) and (c). Here we have applied a local gate $V=0.1$ eV throughout the calculations. In (d), we compare the transmission of all three modes at a fixed Fermi energy $E_F=33$ meV.}
\label{Figure:transmission}
\end{figure}

For the magnetically doped Sb$_2$Te$_3$ QAH system, non-chiral edge modes show up and we consider the co-existence of a pair of non-chiral edge modes (mode I and mode III) and a chiral mode (mode II). Following a similar approach, we make use of the edge state Green function and plot chemical potential and $\theta$ dependence of transmissions $q_i$s for all three modes in Fig.\ref{Figure:transmission} (a) to (c). Here the white dashed line shows the fixed chemical potential $E_F=33$ meV, which we picked to calculate $R_{24}$ in the main text. In Fig.\ref{Figure:transmission} (d), we compare the transmission of all three modes at the fixed Fermi energy $E_F=33$ meV. We notice that transmission of mode II peaks at around $\theta=100$ degree while that of mode I and III peaks at around $\theta=200$ degree. This means that transmission peaks happen at the polarization angle of these edge modes. The above results are consistent with the fact that FM probe is transparent to the electrons with spin parallel to the magnetization of the FM probe.

\section{Magnitude of edge spin polarization for magnetically doped Sb$_2$Te$_3$ QAH systems}
\begin{figure}[t]
\centering
\includegraphics[width=0.6 \textwidth]{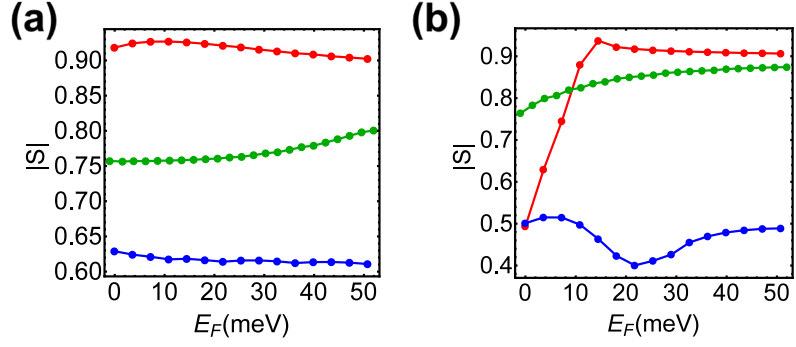}
\caption{Amplitude $|S|$ of spin polarization for magnetically doped Sb$_2$Te$_3$ QAH systems with $V=0$ eV and $V=0.1$ eV are shown in (a) and (b). $|S|$ of edge mode I, II and III are plotted in red, blue and green.}
\label{Figure:spin polarization amplitude}
\end{figure}

We have found that for magnetically doped Sb$_2$Te$_3$ QAH systems, spin polarization ${\bf S}$ is a two dimensional vector living in $y$-$z$ plane, due to the absence of its $x$ component. Therefore, ${\bf S}$ can be written in terms of polar coordinates ${\bf S}=(|S|\cos \theta, |S|\sin \theta)$. In the main text, we have plotted the polarization angle $\theta$ of mode I, II and III. It is shown that mode I (left mover) and mode III (right mover) share a very similar $\theta$ characteristics, and we claim that these two modes should be connected to each other and become trivialized. To make our results complete, here we show that magnitude plot of spin polarization in Fig. \ref{Figure:spin polarization amplitude}. Following the same color convention, $|S|$ of edge mode I, II and III are plotted in red, blue and green. In fact, the value of $|S|$ can be closely related to the calculation details. For example, when locating the CEMs in the iterative Green function (DOS) calculation, a small error in locating the peak position of DOS may lead to a different $|S|$ value of the corresponding CEM. Compared to $|S|$, polarization angle $\theta$ is barely affected by such calculation error. Therefore, we would like to point out that polarization angle $\theta$ is more essential in our discussion here.

\section{Decay length of chiral edge states}
\begin{figure}[t]
\centering
\includegraphics[width=0.6\textwidth]{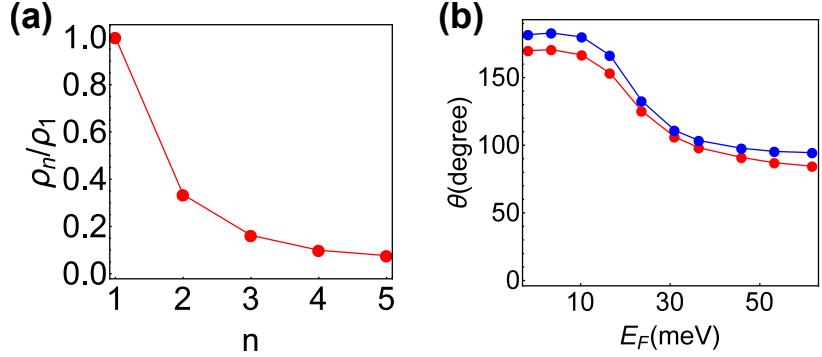}
\caption{In (a), the density of states of the $n$th lattice sites $\rho_n$ is plotted in units of $\rho_1$, for $n=1,2,3,4,5$. It is easy to see that the decay length is simply one lattice constant. In (b), we compare the spin angle $\theta$ of the $n=1$ lattice sites (red line) and that of the entire contribution of $n=1,2,3,4,5$ lattice sites (blue line).}
\label{Figure:decay}
\end{figure}
Let us denote the density of state calculated from the $n$th lattice sites as $\rho_n$, with $n=1,2,3,...$. In Fig.\ref{Figure:decay} (a), we have plotted the value of $\rho_n$ in units of $\rho_1$ of mode II in the Sb$_2$Te$_3$ based QAHE system. It clearly shows that the edge state is highly localized on the edge, and exponentially decays into the bulk. By doing an exponential fit, we find that the decay length is $1.03$ times lattice constant. In our calculation for SP, we have included the total contribution from all $n=1,2,3,4,5$ lattice sites, which we identify as the major contribution. In general one can integrate over the entire semi-infinite space, which is rather costly in calculation, while giving only a small modification to our present results. Actually, we can even show the SP angle $\theta$ is dominated by the contribution from the $n=1$ lattice sites. As shown in Fig. \ref{Figure:decay} (b), we calculated the SP angle $\theta$ of CEM (mode II) with $V_{gate}=0.1$ eV for only $n=1$, and plot $\theta$ evolving with Fermi energy (red line). In comparison, we also plot the SP angle calculated for all $n=1,2,3,4,5$ lattice sites with blue line. It is clear that the $n=1$ lattice sites give the major contributions to $\theta$, while $n=2,3,4,5$ lattice sites together only give a rather small correction to $\theta$. We also would like to emphasize that we are not interested the exact value of CEM SP, which depends on the details of samples in experiments. The essential physics here is that the CEM SP can be manipulated electrically, despite its exact value.

\section{Influence of FM probe on the QAHE}
\begin{figure}[t]
\centering
\includegraphics[width=0.6\textwidth]{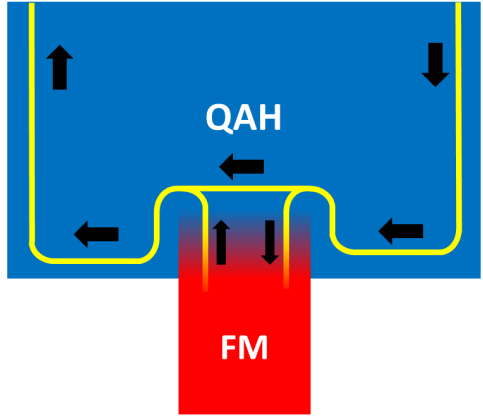}
\caption{We plotted possible reconstruction of chiral edge states under the influence of FM probe. The yellow lines are the chiral edge states. When FM probe destroy the QAHE locally, a new boundary emerges and edge states appear on this new boundary.}
\label{Figure:Influence of FM}
\end{figure}
It is possible that attaching a FM gate to a QAH sample may change the QAH sample locally. For example, the value of CEM spin polarization may be affected in this regime. However, what is physically interesting is the tunability of spin polarization as a result of electrical approaches (chemical potential and local gate tuning). The exact value of CEM spin polarization is not that interesting to us. Therefore, we expect the angle-dependent transverse resistance should still be observed in this case.

In an extreme situation, QAH state is destroyed locally in the vicinity of FM probe, as shown in Fig. \ref{Figure:Influence of FM}. In this case, one can simply treat this local part of QAH sample as an extension of the FM gate. Then a new boundary between FM probe (including its extension) and the QAH sample emerges and edge states still exist on this new boundary, as long as the rest of sample remains in the QAH phase. As a result, our physical picture of this transport experiment still holds.

\section{Stability of edge spin polarization under disorder}
\begin{figure}[t]
\centering
\includegraphics[width=1\textwidth]{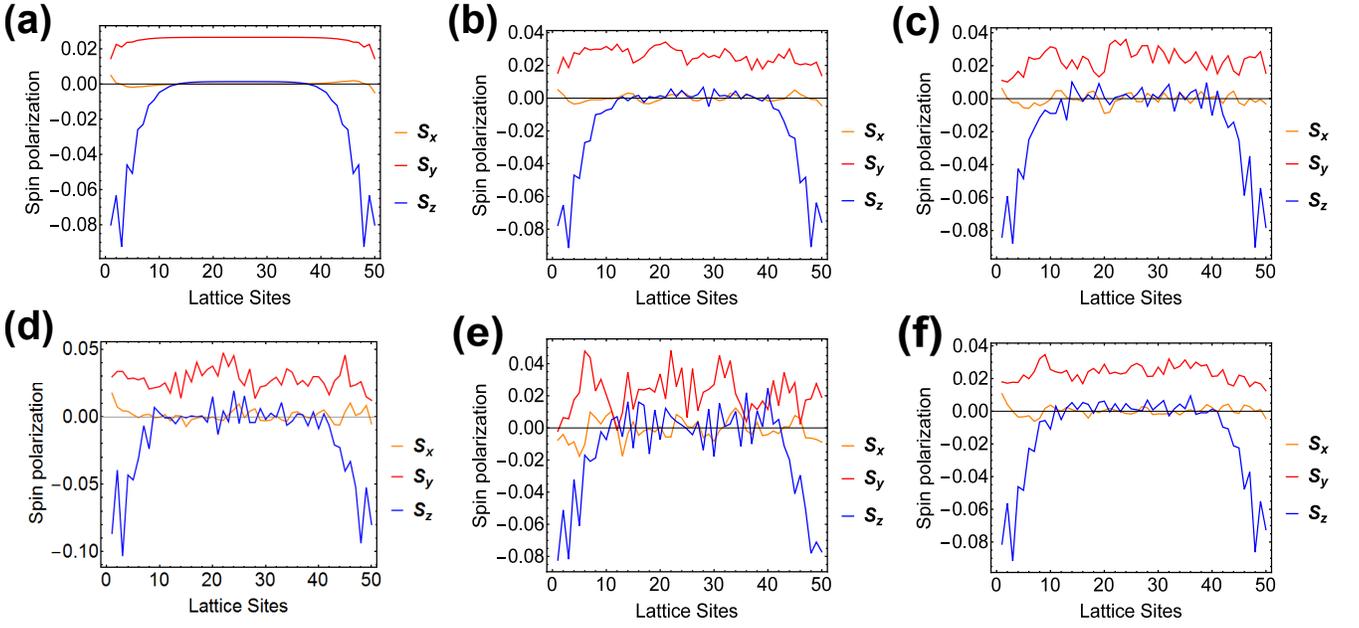}
\caption{Spin polarization of effective QAH model with various disorder strengths $W$ and $N_{ave}=100$ are plotted: (a) $W=0$ eV, (b) $W=0.1$ eV, (c) $W=0.2$ eV, (d) $W=0.3$ eV, (e) $W=0.4$ eV. In (f), we increases $N_{ave}$ to $1000$ with disorder strength $W=0.4$ eV. }
\label{Figure:disorder}
\end{figure}

In this section, we show that the above spin polarization of QAH edge states in the simple effective model (Eq. [1] in the main text) is robust under disorder effect. We start from a finite size real space Green function with a $N_x\times N_y$ lattice configuration with $N_x=50$ and $N_y=2$. To discuss its chiral edge states, we perform iteration of Green function along the $y$ direction, so that system is finite along the $x$ direction while semi-infinite along $y$ direction. In this case, SDOS (spin polarization) of edge states on the $y$-edge will be obtained. Disorder potentials are white-noise-like on-site potentials, and are included locally on the edge for simplicity. The magnitude of disorder potential is between $-W$ and $W$, where $W$ is the disorder strength. We also perform the configuration averaging, the number of which is defined as $N_{ave}$, for disorders.

\begin{figure}[t]
\centering
\includegraphics[width=0.56\textwidth]{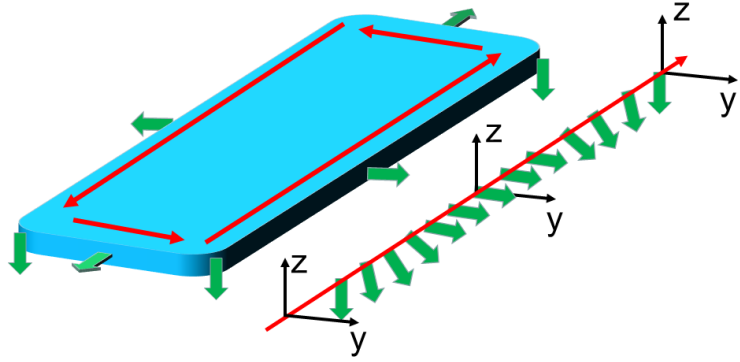}
\caption{In this figure, we give a systematic plot of the edge spin polarization of a finite size QAH sample based on Fig. \ref{Figure:disorder} (a). Red arrow represents edge currents that flow around the QAH sample, while green arrows represent the corresponding spin polarization vectors of the edge current. On the right hand side of the QAH sample, we show in details how the SP of a chiral edge current on the $y$ edge evolves due to the finite size effect.}
\label{Figure:disorder_conf}
\end{figure}

In Fig. \ref{Figure:disorder} (a), we first calculate edge spin polarization without any disorder ($W=0$). A finite local gate voltage ($V_{gate}=0.1$ eV) and a finite Fermi energy ($E_F=0.1$ eV) are considered to guarantee non-zero spin polarization. The $x$-axis represents lattice sites of the one-dimensional lattice chain on the $y$-edge and $y$-axis is for spin polarization. We can clearly see strong finite size effect on the 1D edge chain: At both ends of this 1D chain, there are abrupt changes of spin polarization in all three components. Such a finite size effect can be removed by imposing periodic boundary condition along $x$ direction, and resulting spin polarization is exactly shown in Fig. [1] (a) and (b) (in the main text). This means that for a finite size QAH sample, spin polarization of an electron on the $y$-edge is along $+y$ direction. As shown in Fig. \ref{Figure:disorder_conf}, while approaching the sample corner between $y$-edge and $x$-edge, its spin polarization (green arrows) gradually becomes along $-z$ direction. Finally, when this electron enters $x$-edge, its spin polarization becomes along $-x$ direction.

When disorder is introduced, we can see that the fluctuation of spin polarization increases as the disorder strength $W$ increases, as shown in Fig. \ref{Figure:disorder} (b)-(e). But this disorder induced fluctuation decreases as we increases the configuration averaging number $N_{ave}$ (Fig. \ref{Figure:disorder} (f)). This concludes that spin polarization of chiral edge state in QAHE is robust against disorder effects.

\end{document}